\documentclass[12pt]{spieman}  

\usepackage{amsmath,amsfonts,amssymb}
\usepackage{setspace}
\usepackage{tocloft}
\usepackage{textgreek}
\usepackage{siunitx}

\usepackage[caption=false,font=normalsize,labelfont=sf,textfont=sf]{subfig}

\usepackage[pdftex]{graphicx}
\graphicspath{{figs/}{Figs/}}

\usepackage{array}

\newcommand{\melec}{\mathrm{e^-}}
\newcommand{\elec}{$\melec$}

\newcommand{\unit}[1]{~\mathrm{#1}}

\title{Importance of charge capture in inter-phase regions during readout of charge-coupled devices}

\author[a,*]{Jesper~Skottfelt}
\author[a]{David~J.~Hall}
\author[a]{Ben~Dryer}
\author[a]{Nathan~Bush}
\author[a,**]{Jason~Gow}
\author[a]{Andrew~Holland}
\affil[a]{Centre for Electronic Imaging, Open University, Walton Hall, Milton Keynes, UK, MK7 6AA}

\cftpagenumbersoff{figure}
\cftpagenumbersoff{table}

\begin{document}

\maketitle

\begin{abstract}
The current understanding of charge transfer dynamics in Charge-Coupled Devices (CCDs) is that charge is moved so quickly from one phase to the next in a clocking sequence and with a density so low that trapping of charge in the inter-phase regions is negligible.
However, new simulation capabilities developed at the Centre for Electronic Imaging, that includes direct input of electron density simulations, has made it possible to investigate this assumption further.

As part of the radiation testing campaign of the Euclid CCD273 devices, data has been obtained using the trap pumping method, that can be used to identify and characterise single defects CCDs.
Combining this data with simulations, we find that trapping during the transfer of charge between phases is indeed necessary in order to explain the results of the data analysis. 
This result could influence not only trap pumping theory and how trap pumping should be performed, but also how a radiation damaged CCD is read out in the most optimal way.
\end{abstract}


{\noindent \footnotesize\textbf{*}Address all correspondence to: Jesper Skottfelt, Email: \linkable{jesper.skottfelt@open.ac.uk} }\\
{\noindent \footnotesize\textbf{**}Now at Mullard Space Science Laboratory, University College London, Holmbury St Mary, Dorking, Surrey RH5 6NT, UK }\\

\begin{spacing}{1.1} 

\section{Introduction}
With the ever increasing demand for higher photometric and spatial precision in data coming from space-based observatories, the ability to identify and characterise radiation induced defects in detectors is of high importance. 
Defects are intrinsic in a silicon lattice, even in detectors of a very high quality, and new defects can be created by highly energetic particles, mainly from the Sun, that knock out atoms in the silicon lattice of the Charge-Coupled Device (CCD).
These defects are able to trap electrons during the readout phase of the CCD and release them at a later point in time, which effectively smears the image and thereby negatively affects the image quality. 
To be able to correct for this smearing a high level of knowledge about the trap density and other physical properties is needed. 

While methods such as First Pixel Response (FPR) and Extended Pixel Edge Response (EPER) \cite{Janesick_2001} are only able to give information about average properties of the traps in the CCD, the trap pumping technique \cite{Murray_TrapPumping_2012,Hall_trappumping_2014,Wood_TrapPumping_2014,Hall2017,Wood2017} is able to probe the individual traps.
This means that information about the emission time constant, energy level, capture cross section, sub-pixel positions etc. for the single traps can be extracted and a much better constraint on the trap density can be made. 

The purpose of the VISible imager instrument (VIS)\cite{Cropper_SPIE_2016} on board Euclid\cite{Euclid_Red_book}, the second medium-class mission in the European Space Agency's Cosmic Vision program, is to deliver high resolution shape measurements of galaxies down to very faint limits ($\mathrm{R}\sim25$ at $10\sigma$) in a large part of the sky. The measurement can then be used to infer the distribution of Dark Matter in the Universe. 
However, for this to be possible it is important that the radiation induced traps that accumulates in the detectors over the mission lifetime can be characterised and corrected for to a high precision.
For that purpose trap pumping will be employed as part of the in-orbit calibration routines for the VIS instrument\cite{Cropper_SPIE_2016}. Trap pumping is therefore also a part of the radiation testing campaign, performed at the Centre for Electronic Imaging (CEI), of the CCD273 detectors \cite{Short_2014} that will be used for the VIS instrument. 

In a three-phase device with even-sized phases, as described in Sec.~\ref{sec:3ph}, the trap pumping dynamics can usually be worked out using symmetry considerations, however, for a four-phase device with uneven phase widths, such as the CCD273, this analysis can be more complicated.
We have therefore used the the CEI CCD Charge Transfer Model (C3TM) described in Ref.~\citenum{Jatis2017} (previously named OUMC) to simulate different trap pumping schemes.

It has previously been believed that the transfer of charge between phases is sufficiently quick and over sufficient distance that the combined density/timing prevents trapping in the inter-phase regions. Inter-phase trapping was not observed in preliminary testing and not expected following discussions with relevant experts in the field.
Combining the C3TM simulations with data trap pumping data from a pre-irradiated CCD273 device, we are able to do further investigations into this assumption.

\section{Trap pumping schemes}\label{sec:schemes}
Numerous papers on trap pumping for defect identification have been published over recent years (see for instance Refs.~\citenum{Hall_trappumping_2014} and  \citenum{Hall_SPIE_2016}). This section will therefore only give a short introduction to the trap pumping theory and then detail some of the pumping schemes that can be used. 

\subsection{Three-phase device} \label{sec:3ph}
Trap pumping in a three-phase device is usually done by clocking the charge from phases 1-2-3-1'-3-2-1, where 1' indicates phase 1 in the next pixel.  
The time the charge spends under each phase, the phase time ($t_{ph}$), is the same for all the steps. 

\begin{figure}[ht]
\centering
\includegraphics[width=0.7\linewidth]{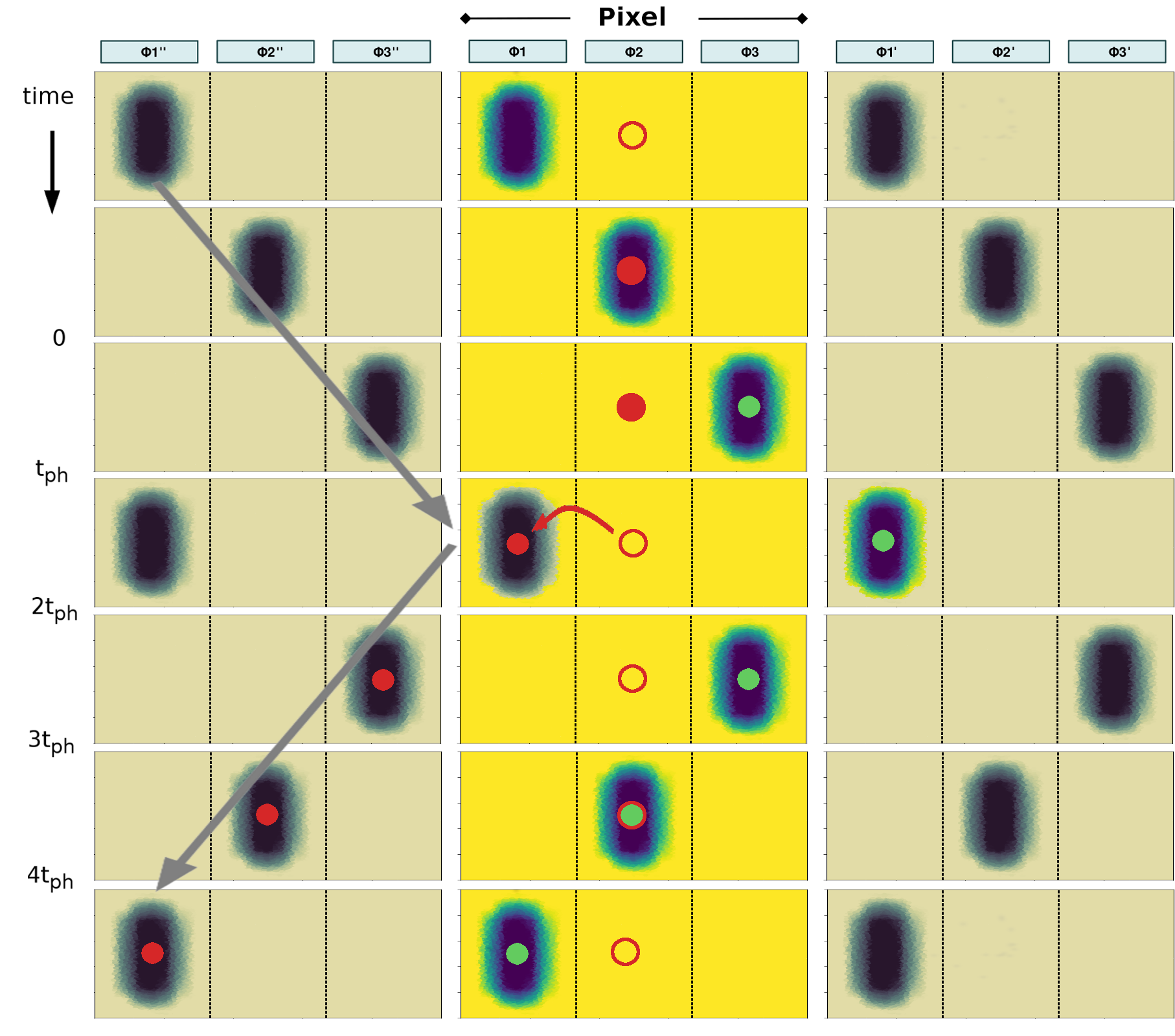}
\caption{Pumping from phases 1, 2, 3, 1', 3, 2 and back to 1, where 1' denotes the first phase in the next pixel. A trap under phase 2 can capture an electron, and if the emission of the electron happens between $t_{ph}$ and $2t_{ph}$ then the electron will be deposited in the adjacent charge packet.
\label{fig:3_phase_tp}}
\end{figure}

\begin{figure}[ht]
\centering
\includegraphics[width=0.6\linewidth]{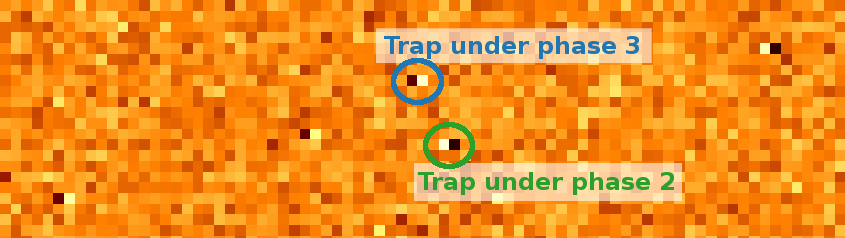}
\caption{Example of dipoles from trap pumped data. The direction of the dipoles reveals under which phase the trap is located. 
\label{fig:dipoles}}
\end{figure}

From symmetry considerations it can be shown that a traps under phase 2 or 3 that captures an electron will release it into the adjacent charge cloud if the emission time constant ($\tau_e$) of the trap is close to $t_{ph}$, as shown in Fig.~\ref{fig:3_phase_tp}.
This creates a dipole, which can be amplified to be distinguishable from shot noise and readout noise (see Fig.~\ref{fig:dipoles}) by repeating the clocking cycle a suitable number of times.

\begin{figure}[ht]
\centering
\includegraphics[width=0.6\linewidth]{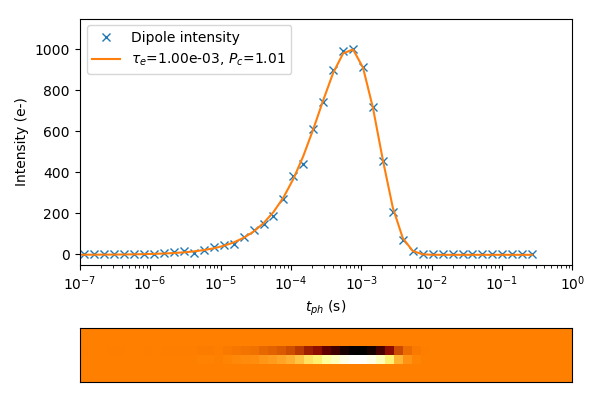}
\caption{(Lower panel) Simulation using C3TM of a trap pumped over a range of $t_{ph}$ values, producing dipoles of different intensities. (Upper panel) Intensity of dipoles in lower panel is plotted on a matching x-axis (crosses) and fitted with Eq.~\ref{eq:tph} (fit shown as solid line), thereby obtaining $\tau_e$ and $P_c$ for the given trap.
\label{fig:single-dipole-fit}}
\end{figure}

By pumping with a range of $t_{ph}$ values it is possible to determine the emission time constant $\tau_e$ for the trap by fitting the dipole intensity curve, shown in Fig.~\ref{fig:single-dipole-fit}, with
\begin{equation}
I_{12} = N \cdot P_c \cdot \left[\exp\left(\frac{-t_{ph}}{\tau_e}\right) - \exp\left(\frac{-2t_{ph}}{\tau_e}\right) \right]\quad, \label{eq:tph}
\end{equation}
where $N$ is the number of cycles and $P_c$ is analogue to the probability of capture. 

If the whole process is done at multiple temperatures even more information about the trap, such as the energy level and emission cross section, can be retrieved. 

\subsubsection{Position of trap in pixel}

A trap under phase 2 will move charge in the opposite direction of a trap under phase 3, so the polarity of the polarity of the resulting dipole will also be opposite and this thus reveals under which of the two phases the trap is. 
A trap under phase 1 will in this situation be filled with charge from two charge packets and the dipole will never form. To detect traps under phase 1, the pumping cycle therefore needs to be started under phase 2 or 3. 

C3TM\cite{Jatis2017} is a Monte Carlo model that simulates the physical processes taking place when transferring signal through a radiation-damaged CCD. 
The software is based on Shockley-Read-Hall theory, and is made to mimic the physical properties in the CCD as closely as possible. The code runs on a single electrode level and takes three dimensional trap position, potential structure of the pixel, and takes device specific simulations of electron density as a direct input, thereby avoiding the need to make any analytical assumptions about the size and density of the charge cloud.

With C3TM it is therefore possible to make a map of the dipoles produced by a traps depending on their position in the pixel. 
This is done by putting traps with a known $\tau_e$ value at a range of sub pixel positions in a single line of pixels such that the first trap is placed in pixel 5 and at subpixel position .005, the next at 10.015, then at 15.015, etc. This means that 100 traps, all with the same $\tau_e$ value, will have sub-pixels positions spread evenly over the pixel. 
A trap pumping scheme is then simulated for this line for a range of $t_{ph}$ values, thus creating a dipole curve for each trap similar to the curve shown in the lower panel of Fig.~\ref{fig:single-dipole-fit}.

\begin{figure}[ht]
\centering
\includegraphics[width=0.7\linewidth]{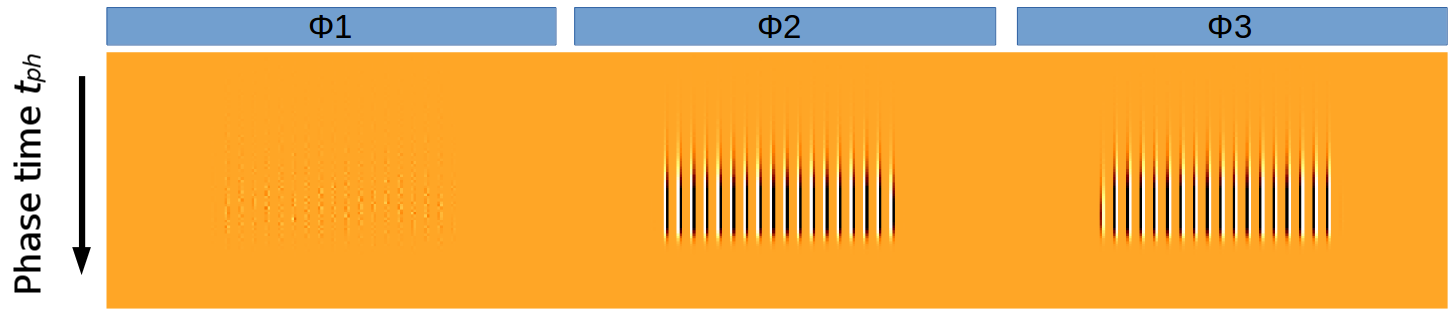}
\caption{Dipole map of a three-phase device using the standard 1-2-3-1'-3-2-1 clocking scheme, showing that traps under phases 2 and 3 will pump in different directions, and that traps under phase 1 will not pump. A signal level of 100,000~\elec{} is used, and the extend of where trap pumping occurs under phase 2 and 3 therefore gives and idea of the width of the charge cloud. 
\label{fig:3ph-dipolemap}}
\end{figure}

For a three-phase device using the standard clocking scheme the map of dipoles are shown in Fig.~\ref{fig:3ph-dipolemap}. As expected this shows that a trap under phase 2 will pump electrons into the charge cloud in the pixel to the left, and a trap under phase 3 will pump to the pixel to the right, while traps under phase 1 does not pump.

\subsubsection{Capture probability}
The probability of capture $P_c$ is in Shockley-Read-Hall theory \cite{Shockley_Read_1952,Hall_1952} described as a decay process
\begin{equation}
P_c = 1-\exp\left(\frac{-t}{\tau_c}\right) ,
\end{equation}
where $\tau_c$ is the emission time constant defined as 
\begin{equation}
\tau_c = \frac{1}{\sigma n v_{th}} .
\end{equation}
Here is $\sigma$ is the capture cross section, $n$ is the electron density, and $v_{th}$ is the thermal velocity. 
While $\sigma$ and $v_{th}$ are constants, $n$ depends on a number of variables, such as pixel geometry, position of the trap in the pixel, signal size and clocking scheme.
To get the best physical representation of $n$, the C3TM software uses device specific simulations of the electron density as a direct input. A more detailed description of this can be found in Ref.~\citenum{Jatis2017}, but Fig.~\ref{fig:3_phase_tp} gives an indication of the size of the charge cloud at 100,000~\elec{}.
Figures later in the paper will give an idea of how the size of the charge cloud changes with signal level.

\subsection{Four-phase device}
\begin{figure}[ht]
\centering
\includegraphics[width=0.7\linewidth]{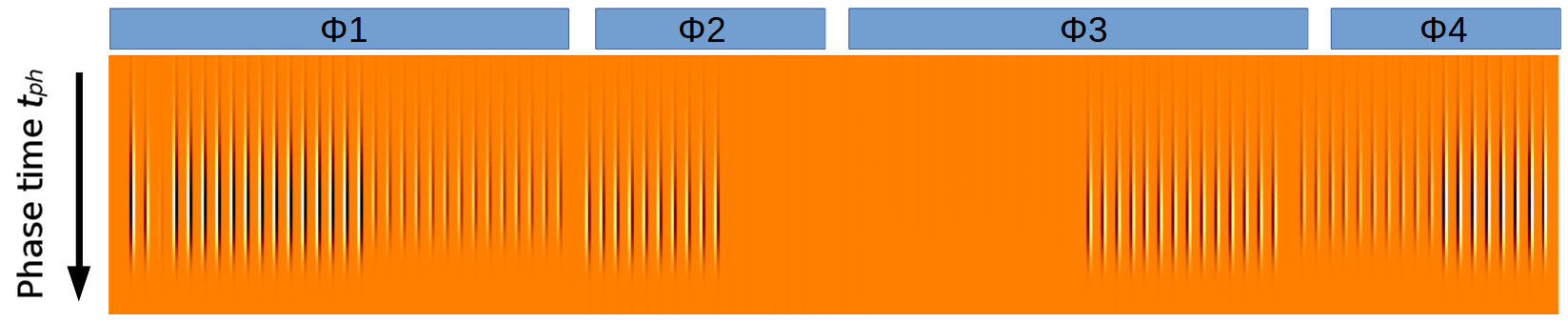}
\caption{Simulation of a four phase non-overlapping trap pumping scheme (1-2-3-4-1'-4-3-2-1) at 100,000~\elec{}, showing that dipole intensities needs to be fitted with three different equations (Eqs.~\ref{eq:tph}, \ref{eq:tph23} and \ref{eq:tph14}) depending on the position of the trap in the pixel.
\label{fig:4ph_novl}}
\end{figure}

The four phase structure of the CCD273's parallel (or image) register means that there are several possible trap pumping clocking schemes to consider. 
An obvious choice would be to simply add the extra phase and pump 1-2-3-4-1'-4-3-2-1 as presented in Ref.~\citenum{Hall_SPIE_2016}. The extra phase step means that depending on under which phase, and even where in the phase, the traps are positioned, the dipole intensity needs to be fitted with either Eq.~\ref{eq:tph} or one of the two following equations
\begin{eqnarray}
I_{23} &=& N \cdot P_c \cdot \left[\exp\left(\frac{-2t_{ph}}{\tau_e}\right) - \exp\left(\frac{-3t_{ph}}{\tau_e}\right) \right] \label{eq:tph23}    \\  
I_{14} &=& N \cdot P_c \cdot \left[\exp\left(\frac{-t_{ph}}{\tau_e}\right) - \exp\left(\frac{-4t_{ph}}{\tau_e}\right) \right]  \, .  \label{eq:tph14}
\end{eqnarray}
This means that three different $\tau_e$ values could be found for the same trap. By starting at multiple starting points for the trap pumping cycle, it would be possible to get information relating not only to which phase the trap is under, but also where in the phase. 
However, with the 4-2-4-2 \textmu m widths of the phases in the CCD273, this information becomes very difficult to disentangle. An example of this is shown in Fig.~\ref{fig:4ph_novl}, where C3TM is used to simulate how the dipole of a trap located at various points in a pixel would look.

\begin{figure}[ht]
\centering
\includegraphics[width=0.7\linewidth]{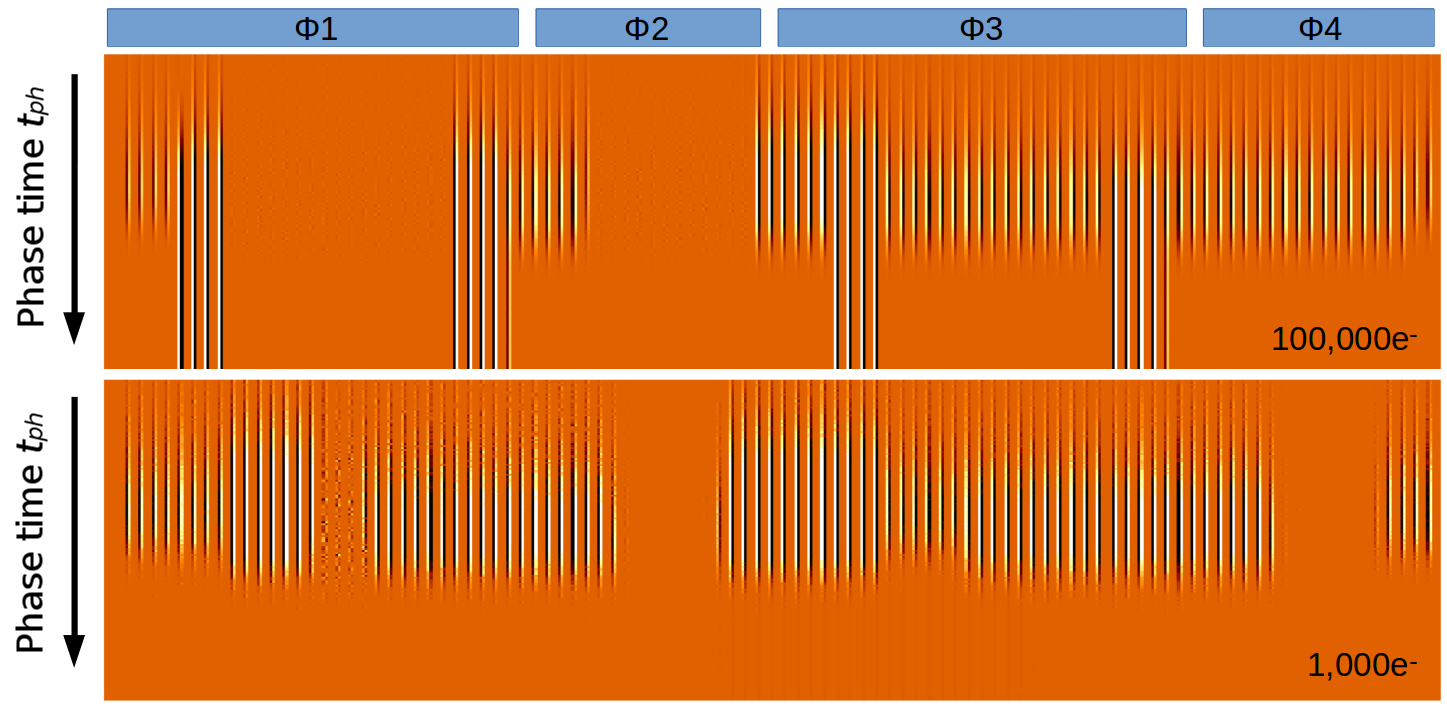}
\caption{Simulation of a four phase overlapping trap pumping scheme (12-23-34-41'-1'2'-41'-34-23-12) at two signal levels, 100,000~\elec{} (top), and 1,000~\elec{} (bottom), showing that dipole intensities needs to be fitted with four different equations (Eqs.~\ref{eq:tph}, \ref{eq:tph23}, \ref{eq:tph14}, and \ref{eq:inv_tph}, depending on the position of the trap in the pixel. 
\label{fig:4ph_ovl}}
\end{figure}

In the normal readout mode for the Euclid VIS instrument the parallel register is clocked with overlapping phases. To mimic this more closely, a trap pumping clocking scheme with overlapping phases could therefore also be used, such as 12-23-34-41'-1'2'-41'-34-23-12. A simulation of this pumping scheme is shown in Fig.~\ref{fig:4ph_ovl} for two different signal levels, 100,000~\elec{} and 1,000~\elec{}. 

For the 100,000~\elec{} level it is seen that at certain places the traps will continue to pump for all $t_{ph}$ values over a certain point.
This comes from the uneven sizes of the phases that means that when the signal is moved from one phase to the next, there is a chance that the neighbouring charge cloud will always be closer to that particular trap, unless the charge is actually in the phase itself. This means that as long as the $t_{ph}$ value is longer than the capture time constant $\tau_c$, then that particular trap will always pump, and we therefore refer to them as `always pump' traps.
The resulting dipole can be fitted with
\begin{equation}
I_{\mathrm{ap}} = N \cdot P_c \cdot \left[1 - \exp\left(\frac{-k \cdot t_{ph}}{\tau_e}\right) \right]\quad, \label{eq:inv_tph}
\end{equation}
where $k$ depends on the pumping scheme and the position of the trap in the pixel.
Fig.~\ref{fig:4ph_ovl} shows that there is a large difference between which types of dipoles we see at different signal levels, and it would therefore be difficult to figure out which equation to use to find the right $\tau_e$ value, if this pumping scheme was used.

Further a three-level clocking scheme\cite{Murray_MultiLevel_2013} is used in normal operations of the Euclid VIS instrument, however, we found that trap pumping with a three-level trap pumping scheme only increased the complexity, and thus makes it even harder to disentangle the single traps.

\subsection{Pseudo-three-phase clocking} \label{sec:ps3ph}
\begin{figure}[ht]
\centering
\includegraphics[width=0.8\linewidth]{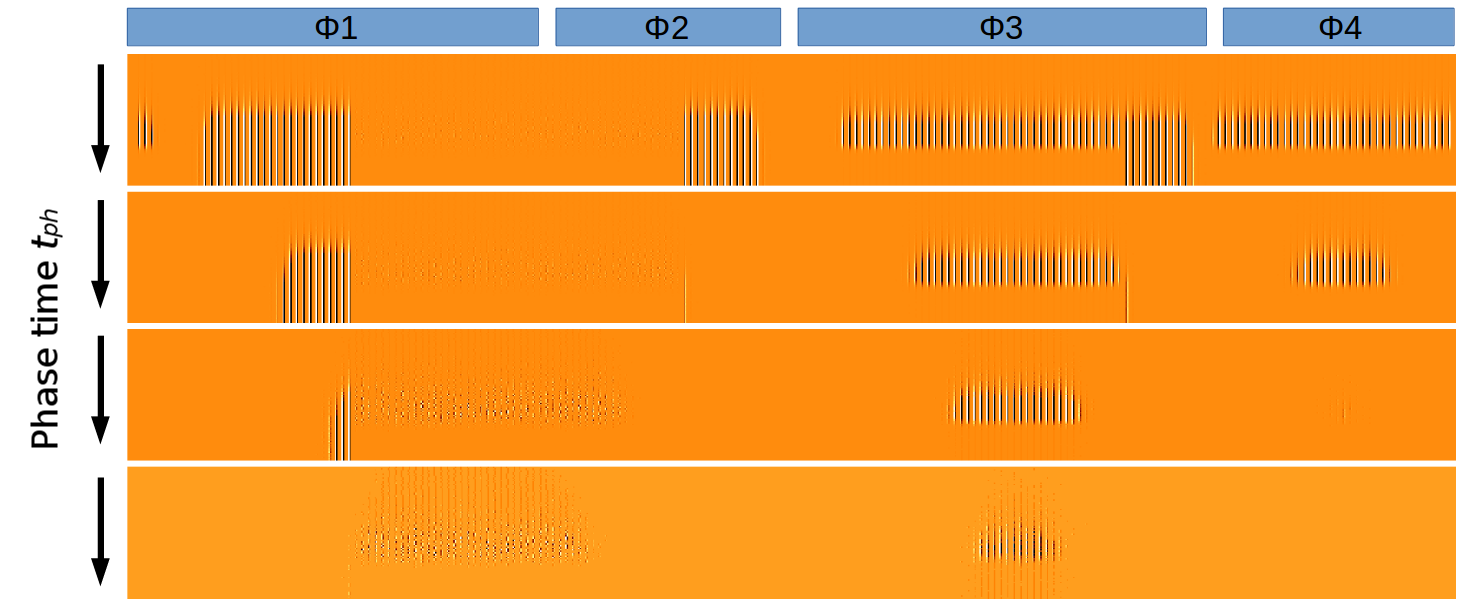}
\caption{Simulation of the pseudo-three-phase clocking scheme (12-3-4-1'2'-4-3-12) at four signal levels; 100,000~\elec, 10,000~\elec, 1,000~\elec, and 100~\elec (from top to bottom).
\label{fig:ps3ph}}
\end{figure}

Another option for a four-phase device is to mimic a three-phase device by coupling two phases together, i.e. 12-3-4-1'2'-4-3-12, thus getting a pseudo-three-phase clocking scheme. Simulations of this scheme is shown in Fig.~\ref{fig:ps3ph} for four signal levels ranging from 100,000~\elec{} to 100~\elec{}, and these show that this scheme should only contain $I_{12}$ dipoles, and `always pump' dipoles. 
It was therefore decided to test if pseudo-three-phase clocking could be used for the Euclid in-orbit calibration routines. 
In order to map the full pixel, and to be able to disentangle the traps, three identical trap pumping schemes are needed, starting at phases 12 (12-3-4-scheme), 23 (23-4-1-scheme), and 34 (34-1-2-scheme). 

\section{Experimental data and initial analysis} \label{sect:initial_analysis}
The data for this analysis have been obtained from the image region of a pre-irradiated CCD273-EM1A device. 
The CCD and its headboard are mounted inside a vacuum chamber that allows the device to be cooled to cryogenic temperatures using a CryoTiger refrigeration system. 
A small LED inside the vacuum chamber delivers a flat-field illumination, however, before any dipole is detected, any pixel-to-pixel non-uniformities or gradients in the flat-field signal is calibrated out. 
A similar setup is used in Ref.~\citenum{Gow_CCD273_2012}, where it is also described in more detail.

Each scheme is run at a number of signal level, ranging from 1,600~\elec{} to 25,000~\elec{}, and three different temperatures (149~K, 153~K, and 157~K), and a range of $t_{ph}$ values between 4~\textmu s to 15~ms. In the following the data taken at 153~k and at a signal level of 1,600~\elec{} are used, but these are representative of the rest of the dataset.

Using an automated dipole detection algorithm, traps over the whole chip is detected and characterised.
The algorithm fits each dipole intensity curve with both Eq.~\ref{eq:tph} ($I_{12}$) and Eq.~\ref{eq:inv_tph} ($I_{\mathrm{ap}}$), and then uses a $\chi^2$ value to determine which $\tau_e$ value to use. Also fitted is a combination of the $I_{12}$ and $I_{ap}$ in order to fit multiple traps in the same pixel, and a constant function, that is used to get rid of false dipoles.
The algorithm outputs the pixel and phase position of the trap, the latter determined by the direction of the dipole, and the best fit values for $\tau_e$ and $P_c$.

\begin{figure}[ht]
\centering
\includegraphics[width=0.8\linewidth]{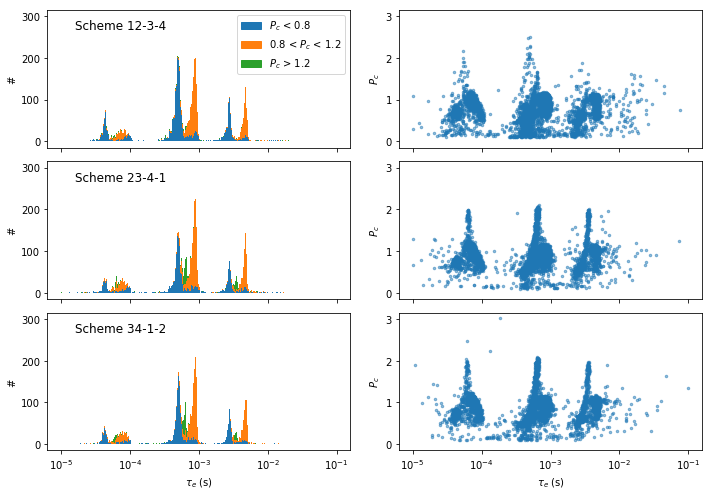}
\caption{$I_{12}$ dipoles from Trap pumping data made for 1600~\elec{} at 153~K. All plots are made from pseudo-three-phase clocking data; 12-3-4-scheme (top), 23-4-1-scheme (middle), and 34-1-2-scheme (bottom). The left-hand side plots are stacked histograms of the $\tau_e$ values, separated into the dipoles where $P_c < 0.8$ (blue), $0.8 < P_c < 1.2$ (orange), and $P_c > 1.2$ (green). The right-hand side plots show the $\tau_e$ value vs. the $P_c$ value for each detected trap. 
\label{fig:1600-novl}}
\end{figure}

Figure~\ref{fig:1600-novl} shows the $I_{12}$ dipoles detected for each of the three schemes. The left-hand side plots are histograms of the $\tau_e$ values, while the right-hand side plots show the $\tau_e$ value vs. the $P_c$ value for each detected trap. 

Common for the histograms for all three schemes are that they have 6 peaks, or rather 3 double peaks, as the peaks two and two are almost a multiple of 2 apart. This suggests that for each double peak only one of them is a 'real' species and the other one is an alias. As it is the rightmost peak for each double peak that is closest to the expected capture probability, $P_c=1$, this suggests that the leftmost peaks that are aliases. 
Double peaks similar to what is found here have been seen in previous trap pumping data studies\cite{Wood2016}, but no conclusive explanation has been found so far. 

Another curios thing is that a number of traps has capture probabilities $P_c > 1$, especially for the 23-4-1 and 34-1-2 schemes, and that these seems to be consistent with the small third peak in between the two main peaks. 

\section{Investigation of aliased peaks}
From the simulation in Fig.~\ref{fig:ps3ph} it is possible to infer under which phase the trap is positioned simply by determining the orientation of the dipole, as described in section~\ref{sec:3ph}. 
The 12-3-4-scheme should therefore pump traps under phase 3 with dipoles in one direction and traps under phase 4 in the other direction. Similarly for the 23-4-1-scheme that will pump phase 4 traps in one direction and phase 1 traps in the other.
This means that traps under phase 4 should be detected by both the 12-3-4-scheme and the 23-4-1-scheme, and thus make it possible to compare the $\tau_e$ value for the same trap from two different schemes.

\begin{figure}[ht]
\centering
\includegraphics[width=0.8\linewidth]{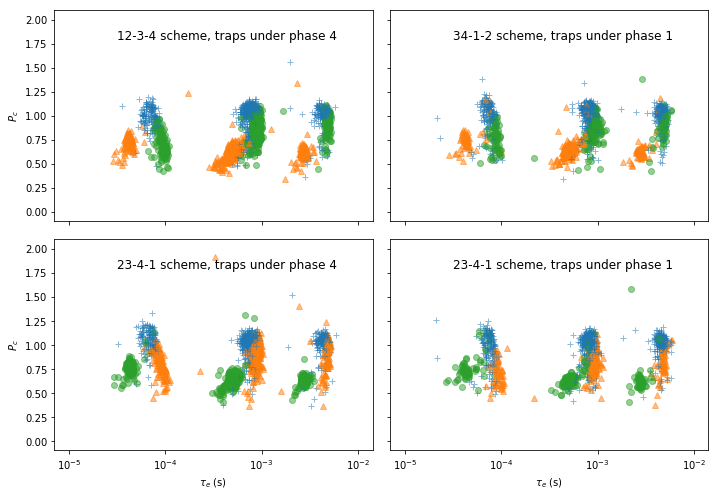}
\caption{\textit{Left}: Comparison of $I_{12}$ dipoles under phase 4 for the 12-3-4-scheme (top) and 23-4-1-scheme (bottom). The blue crosses show the traps where the same $\tau_e$ value within 20\% was found in both schemes. The orange triangles show where the $\tau_e$ value found in the 23-4-1-scheme is more than 20\% larger than the value found in the 12-3-4-scheme, and vice versa for the green circles.
\textit{Right}: Same as left side, but for traps under phase 1, with 34-1-2-scheme (top) and 23-4-1-scheme (bottom).
\label{fig:sch_comp}}
\end{figure}

Figure~\ref{fig:sch_comp} shows the comparison traps under the same phase in terms of $\tau_e$ and $P_c$ values as found by two different schemes. 
The traps found with the same $\tau_e$ value (within 20\%) in the two schemes, are all pumping with the expected $P_c$ value of about 1. Traps where the difference in the $\tau_e$ value is more than 20\% seem to be clustered either at the expected $\tau_e$ and $P_c$ values, or at about $0.6\cdot \tau_e$ and $0.6 \cdot P_c$. 

\begin{figure}[ht]
\centering
\includegraphics[width=0.8\linewidth]{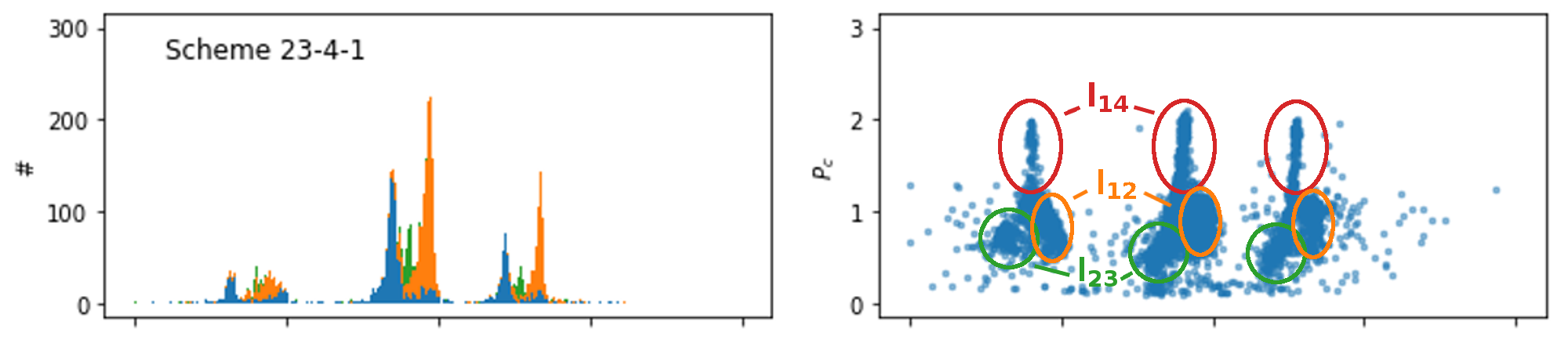}\\
\includegraphics[width=0.8\linewidth]{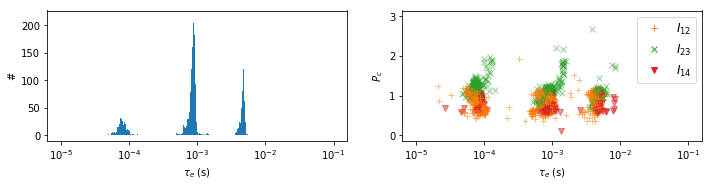}
\caption{\textit{Top}: Cut-out of the 23-4-1-scheme (middle plots in Fig.~\ref{fig:4ph_novl}) with indications of which equations to use for fitting dipoles at different positions in the $\tau_e$ vs. $P_c$ plot.
\textit{Bottom}: Using results from Fig.~\ref{fig:sch_comp} to determine which fit to use for $\tau_e$ and $P_c$ shows that a single peak for each of the three species can be obtained. 
\label{fig:tau_P_eq}}
\end{figure}

Empirically it can be shown that if a distribution of intensities made with $I_{23}$ using $\tau_{e,23}$ and $P_{c,23}$ is fitted with $I_{12}$, then values of $\tau_e \approx 0.6\cdot\tau_{e,23}$ and $P_c \approx 0.6\cdot P_{c,23}$ are found.
Similar if a distribution of intensities made with $I_{14}$ using $\tau_{e,14}$ and $P_{c,14}$ is fitted with $I_{12}$, then values of $\tau_e \approx 0.7\cdot\tau_{e,14}$ and $P_c \approx 1.8\cdot P_{c,14}$ will be found.
We therefore believe that the rightmost peak (shown with the orange circles in top right plot of Fig.~\ref{fig:tau_P_eq}) of each double peak is $I_{12}$ dipoles, while the leftmost peak (green circles) should be fitted with $I_{23}$ instead. The dipoles with $P_c > 1$ (red circles) should be fitted with $I_{14}$ instead.
If the results from Fig.~\ref{fig:sch_comp} are used to determine which equation should be used, then we are able to produce a histogram (bottom left plot of Fig.~\ref{fig:tau_P_eq}) with only one peak per species. 

\begin{figure}[ht]
\centering
\includegraphics[width=0.6\linewidth]{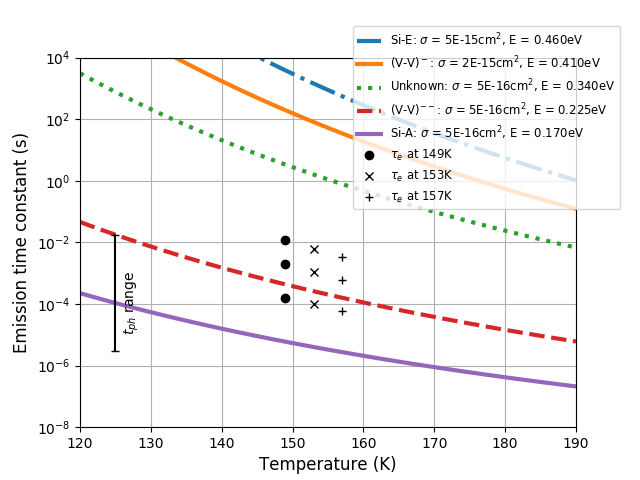}
\caption{Emission time constants of different well-known defects as a function of temperature. The three peaks from from the corrected pseudo-three-phase data is shown as dots, x'es and plusses for 149~K, 153~K and 157~K, respectively. The vertical bar at 125~K shows the range of $t_{ph}$ values used in these trap pumping tests.
\label{fig:temp_vs_taue}}
\end{figure}

By fitting a Gaussian distribution to each of the peaks, the $\tau_e$ value for each of the species can be obtained. 
This has been done for all three temperatures and all the $\tau_e$-peak values are plotted in a $\tau_e$ vs. temperature plot with well-known species in Fig.~\ref{fig:temp_vs_taue}. 
The plot shows that the three peaks follow the same trend as the known species and the energy levels of the new species are estimated to be $0.21~\mathrm{eV}$, $0.24~\mathrm{eV}$, and $0.265~\mathrm{eV}$. 
These energy levels are consistent with Boron-Oxygen and Phosphorus-Carbon impurities \cite{Pichler2004}. Boron and phosphorus are used as dopants in CCDs, and carbon and oxygen are naturally occurring impurities in the silicon wafers from which the CCD are manufactured. 
It should be noted that these energy levels assume a capture cross section of $5\times10^{-16} \unit{cm}^2$ and to get better precision of these two values, the data need to be obtained at a broader range of temperatures.

That none of the known species are found in these data are not surprising, as the data are from an un-irradiated device. However, after irradiation we expect the know species to becomes much more abundant and most likely render the species found here negligible. 
The energy levels found here matches very well with the energy levels found using an alternative pumping scheme, the sub-pixel scheme\cite{PSD11}, that has now been chosen for the Euclid in-orbit calibrations based on the work in this paper. 

\section{Discussion}

It is normally assumed that signal charge moves so fast from one electrode to the next during transfer, and with a low density over the inter-phase spacing, that no charge will be trapped in this transition, i.e. trapping only occurs when the charge sits under the phase during the $t_{ph}$ time. 
This assumption is based on the width of the barrier between the phases, and the speed of an electron moving between changing potentials, and it is only with our new and more advanced models that this can be fully tested.

\begin{figure}[ht]
\centering
\includegraphics[width=0.7\linewidth]{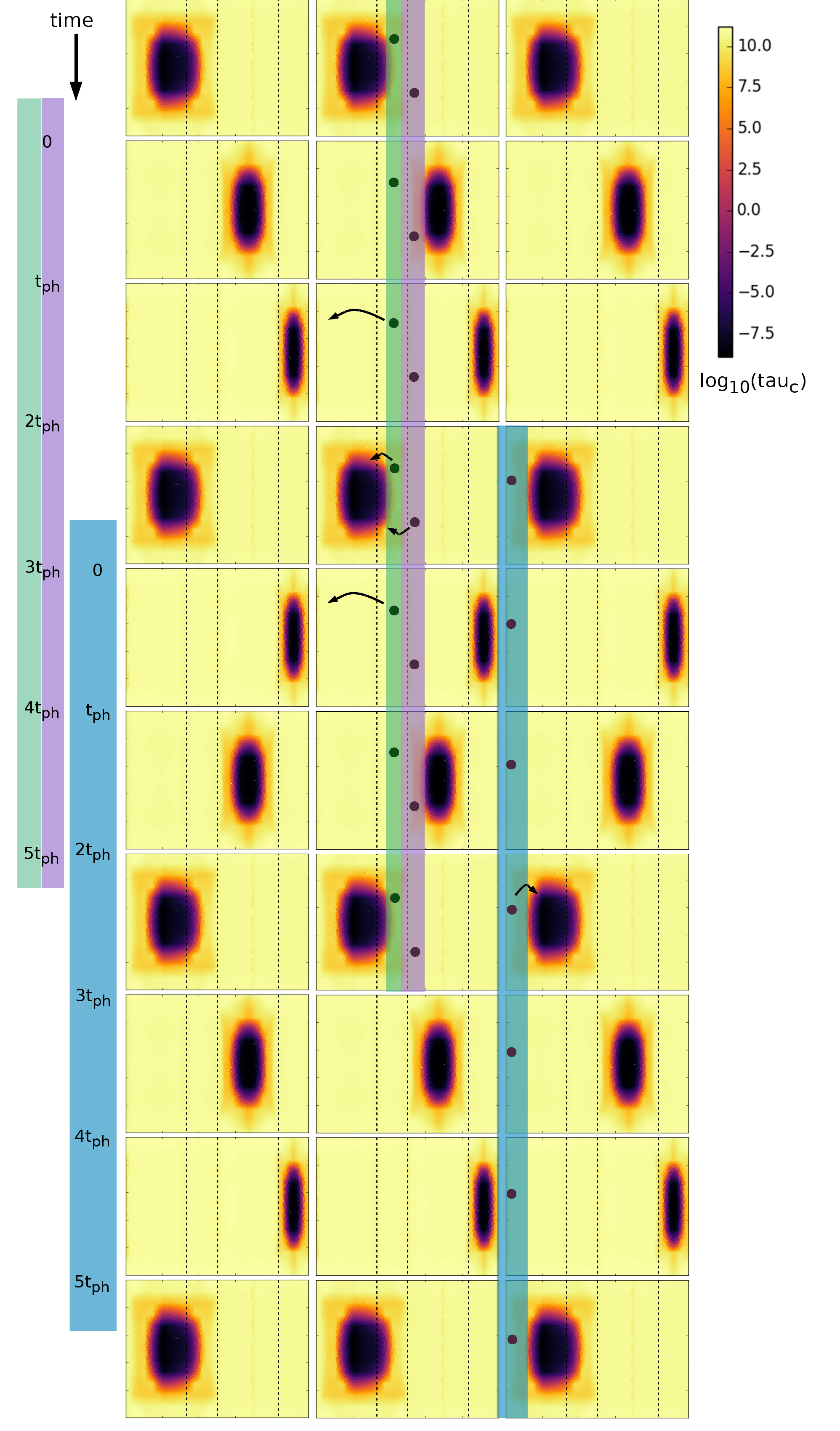}
\caption{Analysis of the movement of the charge cloud during a trap pumping sequence, where the large grey arrows show the movement of a single charge cloud during a single cycle. The black dots show the position of a trap, and the small black arrows shown when a trap would emit to a neighbouring charge from which it has been captured if trapping can happen when charge is moved from one phase step to the next. The purple and blue (green) areas show the position in the pixel of traps creating $I_{23}$ ($I_{14}$) dipoles.
\label{fig:trap_analysis}}
\end{figure}

An analysis of the movement of the charge cloud during the trap pumping sequence with the new simulations, shown in Fig.~\ref{fig:trap_analysis}, shows that if trapping is happening when the charge is moved from one phase step to the next, this would explain the origin of the $I_{23}$ and $I_{14}$ dipoles. 

\begin{figure}[ht]
\centering
\includegraphics[width=0.8\linewidth]{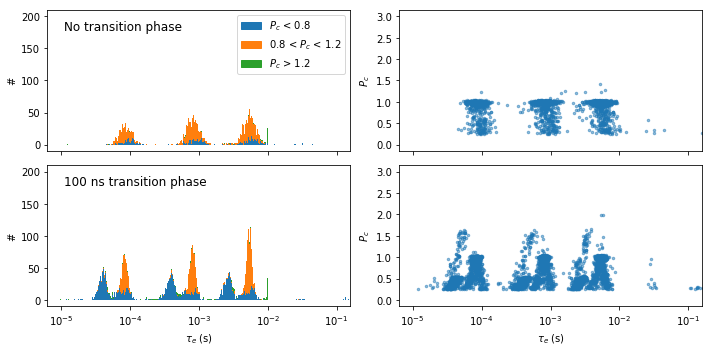}
\caption{Simulation of the pseudo-three-phase scheme using C3TM for 1600~\elec at 153~K, i.e. the same parameters as in Fig.~\ref{fig:1600-novl}. Traps with 5 different energy levels (see details in text) are created at random positions in an array and the trap pumping are simulated without any transition phase (top) and with a 100~ns transition phase (bottom).
\label{fig:ps3ph_sim}}
\end{figure}

A simulation of the pseudo-three-phase scheme on a 2D pixel array was performed using C3TM. Traps with energy levels of $[0.17, 0.21, 0.24, 0.265, 0.34]~\mathrm{eV}$ were inserted in random pixels to reach a trap density for each energy level of $5\times 10^2\unit{traps~cm}^{-3}$, and a signal level of 1600~\elec{} was applied.
When running the simulation without any transition phase (Fig.~\ref{fig:ps3ph_sim} upper panel) the $\tau_e$ vs. $P_c$ does not match the one seen in Fig.~\ref{fig:1600-novl}, however, when a 100~ns transition phase is added (Fig.~\ref{fig:ps3ph_sim} lower panel) they look much more alike. 

\begin{figure}[ht]
\centering
\includegraphics[width=0.7\linewidth]{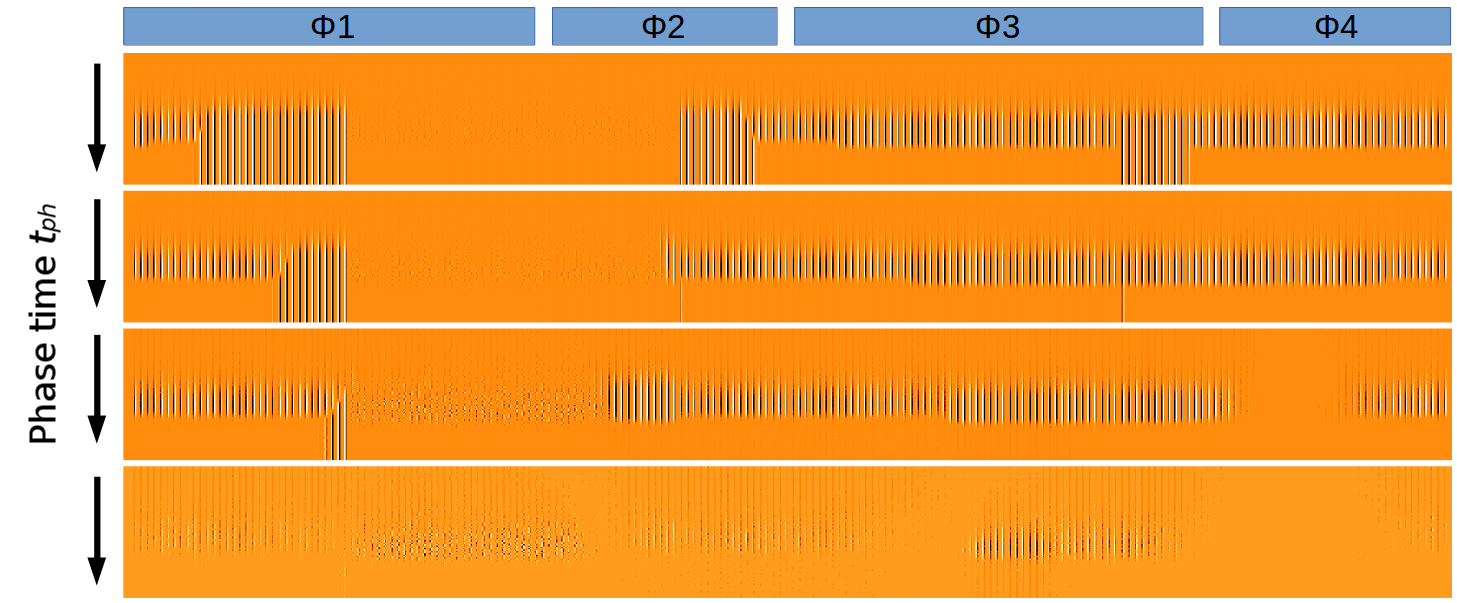}
\caption{Same as Fig.~\ref{fig:ps3ph}, but with a 100~ns transition between each phase step. 
\label{fig:ps3ph_ovl}}
\end{figure}

Figure~\ref{fig:ps3ph_ovl} is a simulation of the 12-3-4-scheme, but with a 100~ns transition step in between each phase step. This shows a much more complicated dipole map than Fig.~\ref{fig:ps3ph} where no transition step is included, and it seems to confirm the positions of the $I_{23}$ and $I_{14}$ dipoles. 
It also shows that the position of the different dipole types are much more dependant on the signal level, than if there is no transition step, and this might make it much harder to disentangle the positions of the traps. 

\begin{figure}[ht]
\centering
\includegraphics[width=0.7\linewidth]{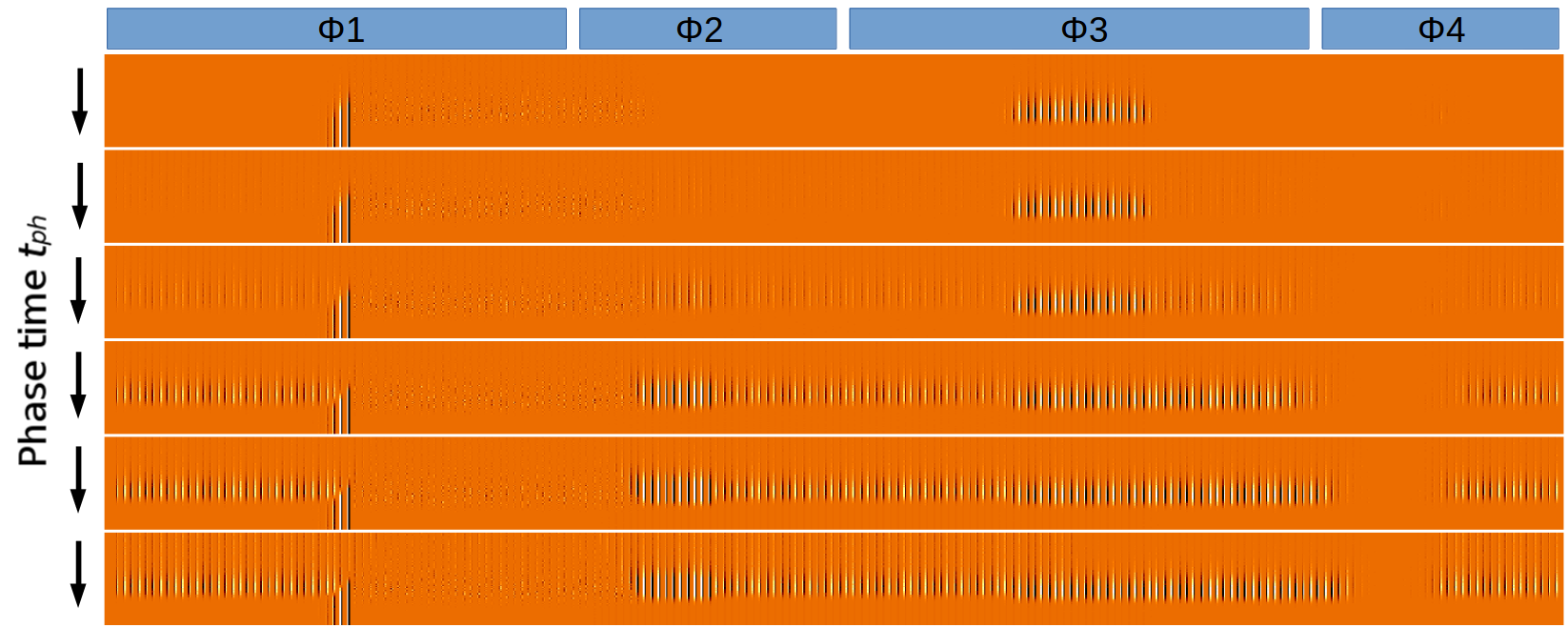}
\caption{Simulation of the 12-3-4-scheme at a signal level of 1,000~\elec, for a transition phase times of; 0~s, 1~ns, 10~ns, 100~ns, 1~\textmu s, 10~\textmu s (from top to bottom).
\label{fig:var-trns}}
\end{figure}

Figure~\ref{fig:var-trns} shows simulations of the 12-3-4-scheme at a range of different transition-step times. This shows that even with a 1~ns transition phase the $I_{23}$ and $I_{14}$ dipoles are visible, although very faint, and that the intensity of these dipoles rise with longer transition times. 
By doing a more thorough comparison of simulations and lab data, it might be possible to further constrain the length of the transition phase. 

Inter-phase trapping seems to have a profound effect on trap pumping theory and it emphasises the importance of being able to simulate the dynamics of the chosen trap pumping scheme to a very high precision. 
However, it could also affect how a radiation damaged CCD is read out in the most optimal way. 
In the case of the Euclid CCD273, the baselined read-out scheme for the parallel register is to always clock with overlapping phases (12-23-34-41'-1'2'-2'3'...). 
This means that traps located between phases will be encountered by the signal charge at each stage of the transfer in the Euclid parallel readout, with the charge held under at least two neighbouring phases at all times. 
If instead a non-overlapping read-out scheme (1-2-3-4-1'-2'-...) had been chosen, inter-phase trapping could potentially occur at each transition step, resulting in worse than anticipated charge transfer efficiency and changing the way in which one might choose to optimise the device readout.

\section{Conclusion}

The common assumption of charge transfer dynamics has been that the charge moves so fast between each step of the sequence, and with a low density over the inter-phase spacing, that no charge will be trapped in this transition. 
However, trap pumping data made as part of the CCD273 irradiation campaign for the Euclid mission does not match with that assumption. 

After a thorough analysis of the data, we find that the only explanation is indeed that inter-phase trapping must be happening. We are further able to identify from which parts of the pixel the inter-phase trapping is most likely to occur. 
These results are backed up by simulations implemented with the CEI CCD Charge Transfer Model (C3TM), that also show that the data are consistent with having a short transition phase between each transfer. The length of the transition phase could be as low as a few ns, but a more thorough comparison of simulations and lab data will be completed in the future in order to constrain this. 

As inter-phase trapping can occur not only when doing trap pumping, but also in a normal read out of the detector, this result can have influence on how a radiation damaged CCD is read out in the most optimal way. 
This will not influence the normal parallel read out of the Euclid CCD273 device, as this is always clocked with overlapping electrodes, but based on this work another trap pumping scheme, the sub-pixel scheme, has been chosen for the Euclid in-orbit calibrations as detailed in Ref.~\citenum{PSD11}.

\bibliography{TP_interphase_trapping} 
\bibliographystyle{spiejour} 

\end{spacing}
\end{document}